\documentclass[apj]{emulateapj}

\usepackage{graphicx}
\usepackage{lineno}

\shorttitle{A minimal width of the arrival direction distribution of UHECRs}
\shortauthors{A.A. Ivanov}

\begin{document}

\title{A minimal width of the arrival direction distribution\\ of ultra-high energy cosmic rays detected with the Yakutsk Array}

\author{A.A. Ivanov}
\affil{Shafer Institute for Cosmophysical Research and Aeronomy,\\
31 Lenin Avenue, Yakutsk 677980, Russia; ivanov@ikfia.ysn.ru}

\begin{abstract}
This paper presents the results of searches for anisotropy in arrival directions of ultra-high energy cosmic rays detected with the Yakutsk Array during the 1974--2008 observational period together with available data from other giant extensive air shower arrays working at present. A method of analysis based on a comparison of the minimal width of distributions in equatorial coordinates is applied. As a result, a hypothesis of isotropy in arrival directions is rejected at the $99.5\%$ significance level. The observed decrease in the minimal width of distribution can be explained by the presence of cosmic ray sources in energy intervals and sky regions according to the recent indications inferred from data of the Yakutsk Array and Telescope Array experiments.
\end{abstract}


%\linenumbers

%%%%%%%%%%%%%%%%%%%%%%%%%%%%%%%%%%%%%%%%%%%%%%%%%%%%%%%%%%%%%%%%%%%%%%%%%%%%%%%%%%%%
\section{Introduction}
The origin of ultra-high energy cosmic rays (UHECRs) is a long-standing challenge in cosmic ray (CR) physics. Extensive air shower (EAS) arrays detecting CRs at energies above 1 EeV (=$10^{18}$ eV) observe mainly isotropic arrival directions with no sign of fluxes from sources significantly exceeding instrumental errors (e.g., \cite{Joint,Kampert}).

At the same time, there are indications of the small size anisotropy in arrival directions revealed by means of a comparison of CR intensities in adjacent sky regions (e.g., \cite{Wavelet,Sommers,TA}). To confirm or refute the indications it is useful to diversify approaches and methods in analysis of CR arrival directions besides independent experimental data.

Recently, a method of the minimal width of arrival direction distribution (MWADD) was used\footnote{for a method in the context of directional statistics see Appendix} to search for the sources of UHECRs in a circle of right ascensions \citep{MinVar}. The method is a specific variant of testing for the equality of variances of populations, as an alternative to the equality of the means \citep{WSRT}.

In this paper, a two-dimensional generalized method is developed for application in equatorial coordinates to test a null hypothesis, $H_0$, of isotropy in the arrival directions of CRs and an alternative hypothesis with fitted position and luminosity of a possible source of CRs. Our aim is to reject or confirm and refine characteristics of the possible CR sources indicated in the previous papers.

The present enhancement of the method consists in consideration of the non-uniformity of the array acceptance area in CR arrival angles due to the unequal time-integrated flux from different directions of the sky.

%%%%%%%%%%%%%%%%%%%%%%%%%%%%%%%%%%%%%%%%%%%%%%%%%%%%%%%%%%%%%%%%%%%%%%%%%%%%%%%%%%%%
\section{The Yakutsk Array experiment and sampling of the data set}
The main purpose of the Yakutsk Array\footnote{Website: http://eas.ysn.ru} is to investigate CRs measuring EAS in the energy range of $10^{15}-10^{20}$ eV. Construction of the array near Yakutsk, Russia, at geographical coordinates $61.7^0N,129.4^0E$, 105 m above sea level (1020 g/cm$^2$), was completed in 1973 \citep{Mono}. During years, the array has been reconfigured several times. Before 1990, the total area covered by detectors was $\sim17$ km$^2$; now, it is $8.2$ km$^2$. At present, it consists of 58 ground-based and four underground scintillation counters to measure charged particles (electrons and muons), and 48 detectors of the air Cherenkov light \citep{Kashiwa,MSU}.

EAS events selection from the background is realized with a two-level trigger of detector signals: The first level is a coincidence of signals from two scintillation counters in a station within 2 $\mu$s; the second level is a coincidence of signals from at least three nearby stations within 40 $\mu$s. Functioning procedures and the types of array detectors are described in \cite{Mono,Malfa,Modernize}.

The location of the shower core is based on the fitting of the particle lateral distribution by the Greisen-type trial function. Core location errors are $\sim30$ to $\sim50$ m depending mainly on the number of triggered stations in the EAS event \citep{Mono}.

Arrival angles of EAS primary particles are calculated in the plane shower front approximation using the trigger times of stations. A clock pulse transmitter at the center of the array provides a pulse timing of $\sim100$ ns accuracy. Errors in arrival angles depend on the primary energy, decreasing from $\sim7^0$ at $E=1$ EeV to $\sim3^0$ above $E=10$ EeV. For more detailed information see \cite{NJP,MSU}.

In contrast to two other giant arrays functioning at present, Pierre Auger Observatory (PAO, \cite{PAOdet}) and the Telescope Array (TA, \cite{TAdet}) with fluorescent detectors, atmospheric Cherenkov light is used here to estimate the energy of primary particles initiating EAS \citep{JETP,NJP,Telescope}.

In this work, the same sample of the Yakutsk Array data is used as in the previous paper \citep{MinVar}, consisting of EAS events detected in the period January 1974 -- June 2008, with shower axes within the array area with zenith angles $\theta<60^0$. The unified energy estimation procedure is applied to all showers throughout the period, following \cite{Kashiwa}. An additional cut is caused by the increased threshold energy, which will be set out in the next subsection.

%%%%%%%%%%%%%%%%%%%%%%%%%%%%%%%%%%%%%%%%%%%%%%%%%%%%%%%%%%%%%%%%%%%%%%%%%%%%%%%%%%%%
\subsection{Exposure of the ground EAS array for celestial regions}
\label{sctn:Exposure}

\begin{figure}
    \centering
    \includegraphics[width=\columnwidth]{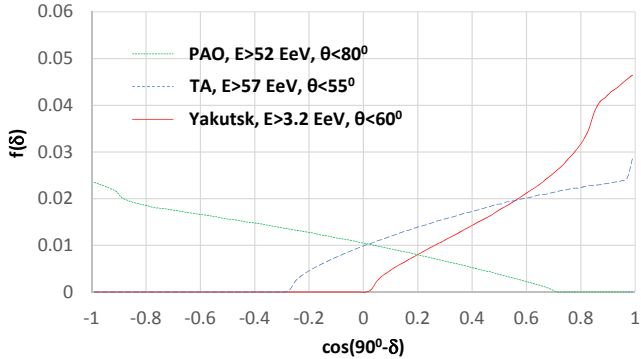}
    \caption{Expected declination distribution of the isotropic cosmic rays detected with EAS arrays. $N=10^6$.}
    \label{Fig:EqFig}
\end{figure}

Because the array aperture is bounded in the zenith angle interval $\theta\in(0,\theta_{thr})$, CRs can be detected only from a part of the sky in a horizontal system. A diurnal cycle of the array functioning provides nearly uniform exposure in right ascensions and apparent non-uniformity in declinations.

A simple and convenient way to calculate the non-uniform exposure of the array is to use a Monte Carlo (MC) method (for basics see, e.g., \cite{MonteCarlo1,MonteCarlo2}; an application to the array exposure is demonstrated by \cite{IzvAN,Wavelet}). An algorithm is based on the angular distribution of isotropic rays in a horizontal system attached to the flat array on the ground. Within the infinitesimal time interval $(t,t+dt)$, with stationary Earth, the azimuthal distribution is uniform and the zenith angle distribution is formed as follows: if ($\theta<\theta_{thr}$) then $f_i(\theta)=\sin(2\theta)/(1-\cos^2\theta_{thr})$, else $f_i(\theta)=0$. If we assume the inverse time $t\rightarrow -t$, so that all the exposed rays move from the array to the sky, then the rays are integrated over the diurnal cycle due to the Earth reverse rotation.

In this algorithm, random directions, $(\phi_i,\theta_i), i=1,...,N$, are sampled in the horizontal system from a uniform distribution in the azimuth and $f_i(\theta)$ in zenith angles. A uniform distribution of the sidereal time is used as well. Then directions are converted, for instance, to equatorial angles $(\phi_i,\theta_i)\rightarrow(\alpha_i,\delta_i)$ (e.g., \cite{SphericalAstronomy1,SphericalAstronomy2}) to form the expected-for-isotropy distribution of CR arrival directions on the celestial sphere.

Resultant declination distributions for the ground arrays working at present are shown in Fig. \ref{Fig:EqFig}, while the right ascension distribution is uniform. Actually, the obtained distributions are formed by the directional exposures of EAS arrays, that is, the effective time-integrated collecting area for a flux from each direction of the sky.

Minor deviations from the uniform distribution in right ascension are caused by the diurnal and seasonal variations of the array exposure. The efficiency of array detection is affected by the weather effects, shutdowns of detectors, the geomagnetic modulation of the EAS event rate, and so on \citep{Weather,Geo,JETP}. Attenuation of showers in the atmosphere results in a deviation from the `isotropic' zenith angle distribution, $f_i$, at low energies. However, above some threshold energy, $E_{thr}$, all these effects can be neglected against statistical errors rising with energy. In this paper, the following values are chosen: $E_{thr}=3.2$ EeV, $\theta_{thr}=60^0$ for the Yakutsk Array \citep{MinVar}, $E_{thr}=52$ EeV, $\theta_{thr}=80^0$ for PAO \citep{PAOdata}, and $E_{thr}=57$ EeV, $\theta_{thr}=55^0$ for TA data \citep{TA}.

\begin{figure}[t]\centering
\includegraphics[width=0.91\columnwidth]{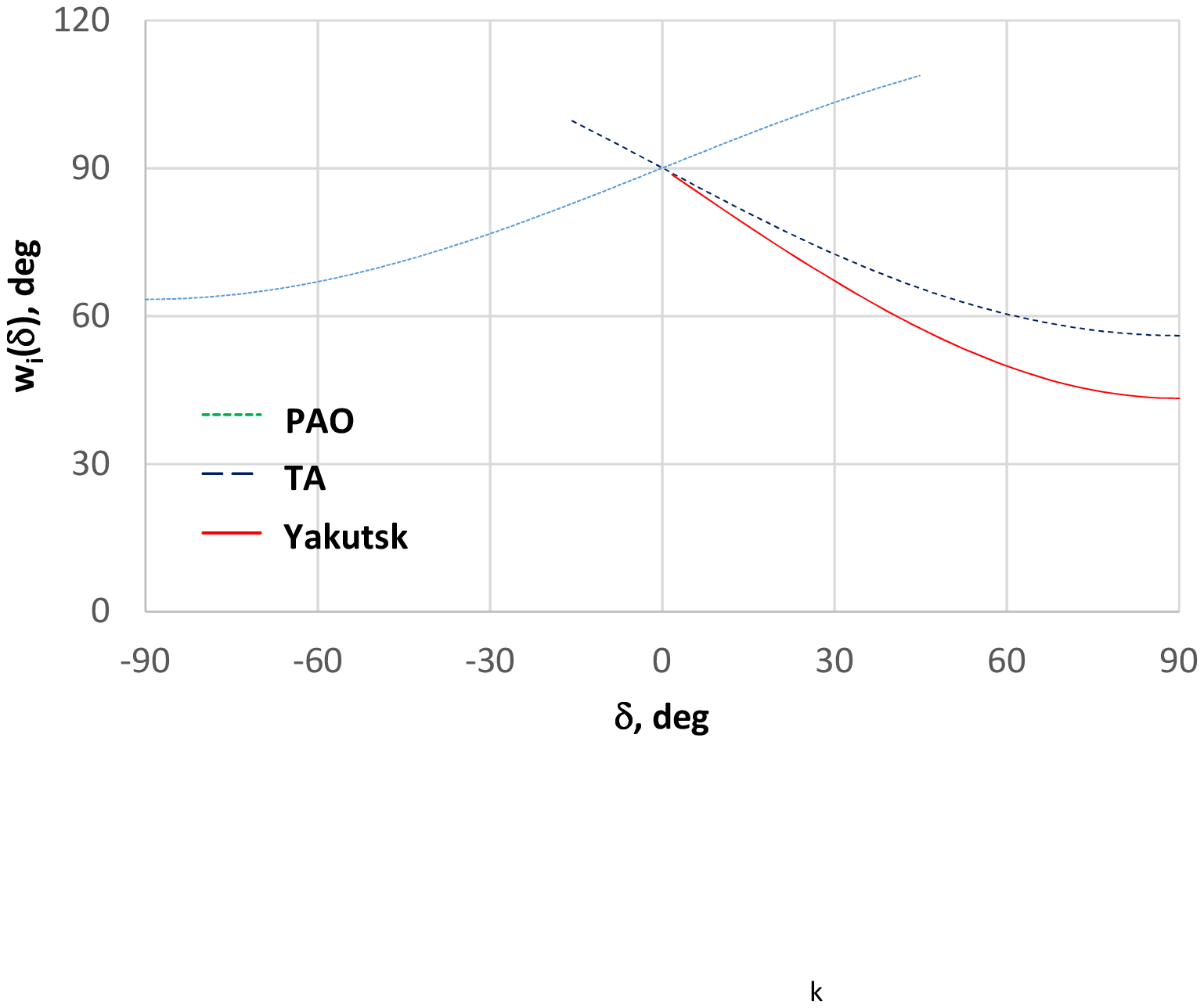}
  \caption{Half width of the isotropic arrival direction distribution as a function of declination. $N=10^6$.}
\label{Fig:AsympW}\end{figure}

%%%%%%%%%%%%%%%%%%%%%%%%%%%%%%%%%%%%%%%%%%%%%%%%%%%%%%%%%%%%%%%%%%%%%%%%%%%%%%%%%%%%
\section{Analysis of CR arrival directions in equatorial coordinates}

\subsection{A method of the minimal width of distribution to analyze arrival directions of CRs}
Isotropic distribution of arrival directions has no mean value because the limits of the region on the sphere that should be integrated over are undefined. Meanwhile, if we assume an arbitrary direction, $(\alpha,\delta)$, as a trial mean, then we can find a dispersion of the isotropic distribution, namely, the width $2\omega_i$, which is independent of the trial mean:
$$
\omega_i=\frac{1}{4\pi}\int_0^{2\pi}d\phi\int_0^\pi \psi\sin\psi d\psi=\frac{\pi}{2},
$$
where $\psi$ is the angular distance to $(\alpha,\delta)$. In the case of $N$ data points on the celestial sphere, a sum of angular distances is applicable instead:
$$
\omega_i=\frac{1}{N}\sum_{i=0}^{N}\psi_i,
$$
where the asymptotic limit is equal to $\pi/2$ as $N$ approaches $\infty$.

So, the width of the isotropic distribution on the sphere is $2\omega_i=180^0$. On the other hand, if there is a source of CRs with the angular size $S\ll\omega_i$ lurking in an isotropic background, then the aggregate width of the distribution: i) reaches the minimum when the trial mean points to the source; ii) has a minimum which is distinctly less than $2\omega_i$, depending on the fraction of the source luminosity in the overall flux of CRs.

For instance, if the flux from the source is half of the total, then $\omega_{min}=45^0$, while for a fraction of 0.1, half of the minimal width is $\omega_{min}=81^0$. It seems that by measuring the width of the distribution of arrival directions, one is able to reject the null hypothesis and to find the coordinates and the fraction of CR flux from a source, if there is any.

%%%%%%%%%%%%%%%%%%%%%%%%%%%%%%%%%%%%%%%%%%%%%%%%%%%%%%%%%%%%%%%%%%%%%%%%%%%%%%%%%%%%
\subsection{Application of the method to Monte Carlo data}
In this section, simulation results of the MWADD method applied to $N$ points on the equatorial sphere taking into account the array exposure are given. In this case, the width of distribution is strongly influenced by the exposure, so the directional dependence $\omega_i(\alpha,\delta)$ should be calculated for the particular array.

It is straightforward to use the MC algorithm described above (Section \ref{sctn:Exposure}) to compute the width with the trial mean scanning the whole $\alpha\in(0,360^0),\delta\in(-90^0,90^0)$ equatorial area. The results for three arrays are shown in Fig. \ref{Fig:AsympW}. At $E>E_{thr}$ the distribution width is the same in right ascensions, so the width variation is shown for the trial mean scanning declinations.

The method is applicable only in searching for a single source, SS, of CRs. Indeed, in the case of two sources located at the angular distance $L$ from each other, with the fractions of CR fluxes $f$ and $(1-f)$, MWADD is $\omega_2=2f(1-f)L$. For opposite sources with equal fractions, a half-width is $\omega_2=\pi/2$, just as in the isotropic alternative. In what follows, we will explicitly suppose an SS of CRs within a particular energy interval.

The statistical power of the MWADD method is the efficiency depending on the sample size $N$. We have to find a lower limit of $N$ needed to reject $H_0$ at a confidence level of 99\% when an alternative hypothesis, $H_1$, is true. To estimate $N_{min}$, we used $H_1$ consisting of SS as a $\delta$-function located in $(\alpha_{SS},\delta_{SS})$, within the field of view of the Yakutsk Array, with the fraction of the total CR flux $f$, and an isotropic background which provides $(1-f)$ of the flux. The distribution width for each trial mean is normalized using the `exposed' isotropic distribution width as a measure.

\begin{figure}[t]\centering
\includegraphics[width=\columnwidth]{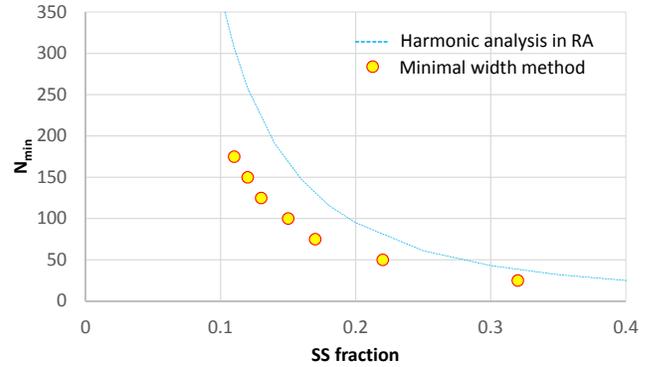}
  \caption{Minimal sample size, $N_{min}$,  needed to reject isotropic distribution if there is a single source, SS, giving a fraction of the total CR flux.}
\label{Fig:Power}\end{figure}

The result of MC simulation is given in Fig. \ref{Fig:Power} in comparison with the power of harmonic analysis in the right ascension\footnote{In other words, the Rayleigh test where the first harmonic amplitude is a measure of dispersion \citep{Jupp}}. The first harmonic amplitude, $A_1$, under $H_1$ is a weighed vector sum of $2$ and $2/\sqrt{N}$ \citep{MinVar}. Using a probability $P(>A_1)=exp(-NA_1^2/4)=0.01$, one can find $N_{min}$ for $H_0$ as a function of $f$. A conclusion to be drawn is that the minimal width method is more powerful than harmonic analysis in R.A.

%%%%%%%%%%%%%%%%%%%%%%%%%%%%%%%%%%%%%%%%%%%%%%%%%%%%%%%%%%%%%%%%%%%%%%%%%%%%%%%%%%%%
\subsection{Application of the method to experimental data}
\subsubsection{Testing the null hypothesis with the Yakutsk Array data}
The Yakutsk Array data at energies above 3.2 EeV are divided into four intervals with the widths $\Delta\lg E=0.25, 0.5$. Below, energy scaling factors for EAS arrays derived from comparison of the observed energy spectra \citep{APJ,CERN,Utah} are used (PAO: 1.04; TA: 0.96; Yakutsk: 0.561). Scaled energy is marked $E_{WG}$.

\begin{figure*}[t]\centering
\includegraphics[width=0.052\textwidth]{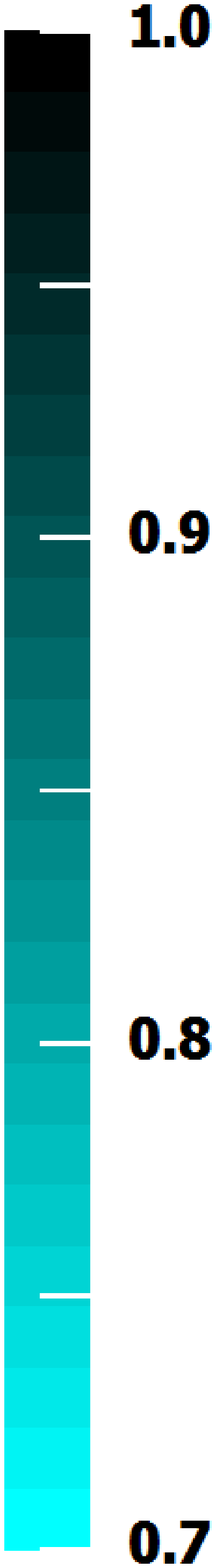}
\includegraphics[width=0.30\textwidth]{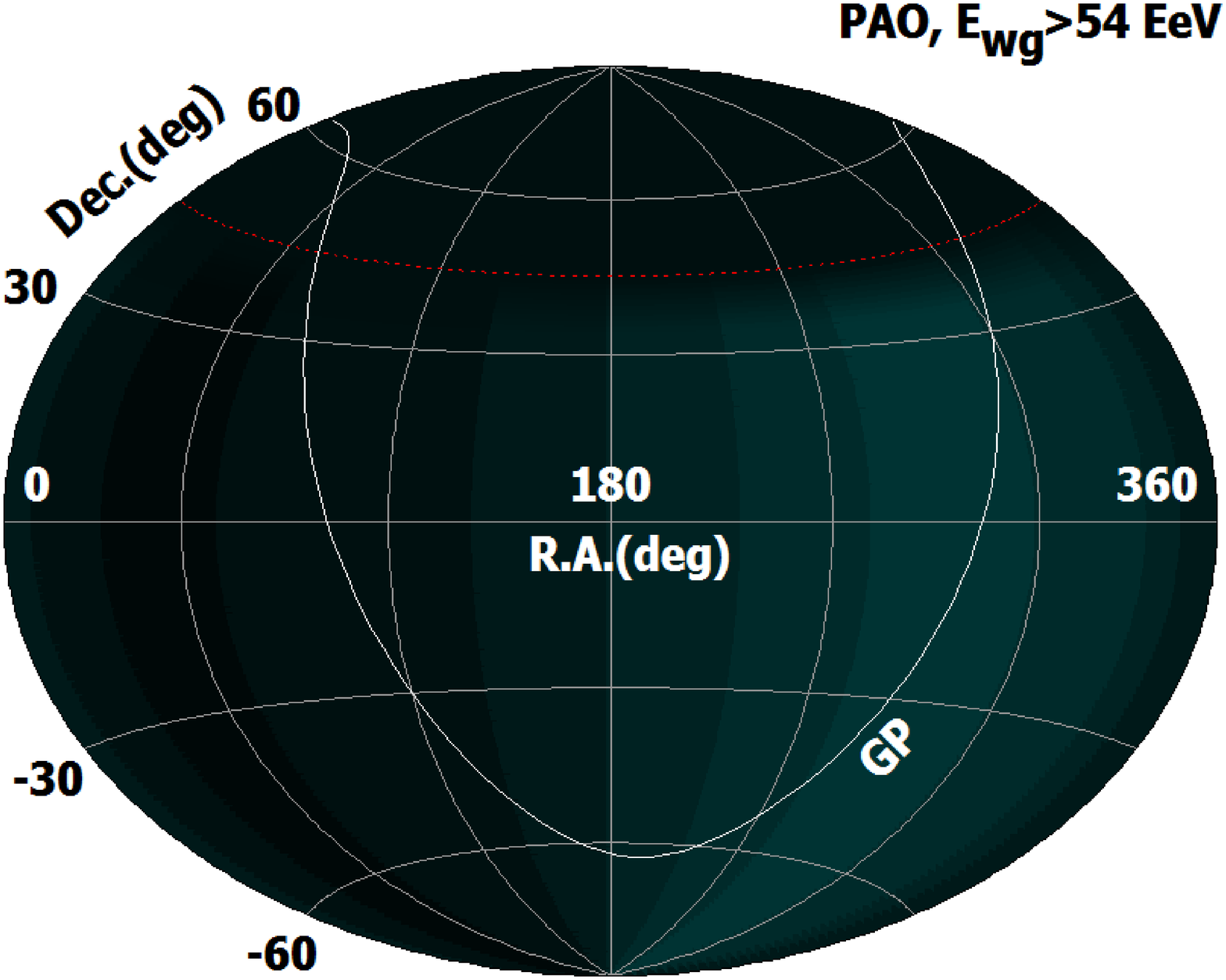}
\includegraphics[width=0.30\textwidth]{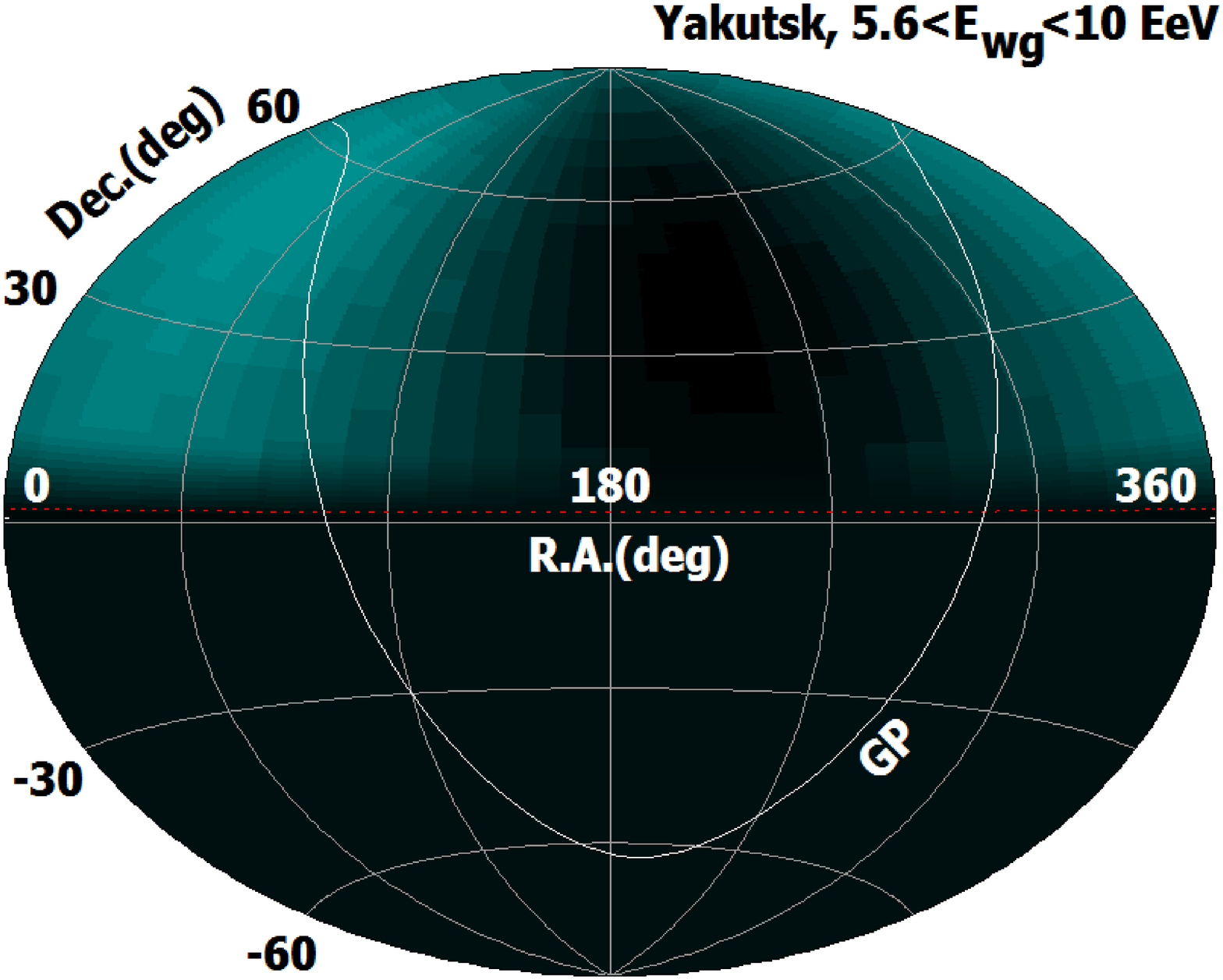}
\includegraphics[width=0.30\textwidth]{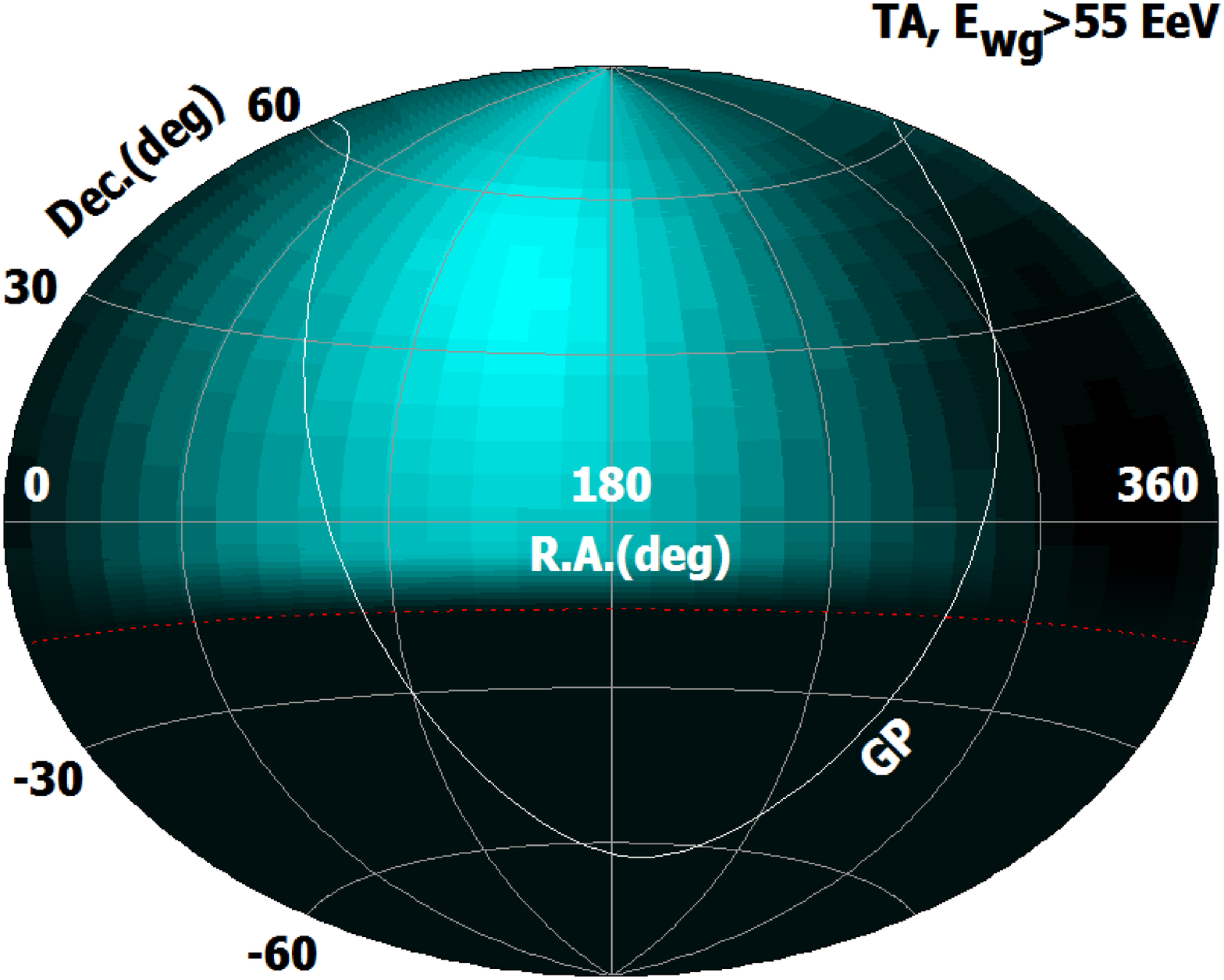}
%  \caption{Ratio of the observed to expected-for-isotropy width of the distribution of arrival directions as a function of the trial mean.}
\caption{Ratio of the observed to expected-for-isotropy width of the distribution of arrival directions in arbitrary colors (a scale bar is given on the left) mapped using Hammer-Aitoff projection of equatorial coordinates. FOV border lines of the arrays are shown by dotted curves.}
\label{Fig:Variance}\end{figure*}

The width of the observed distribution of arrival directions is normalized using the expected isotropic distribution width for a given trial mean. Only in the energy interval $E_{WG}\in(5.6,10)$ EeV there is a definite minimum, $\omega_{obs}/\omega_i=0.89$, of the distribution width of CRs detected with the Yakutsk array, shown in Fig. \ref{Fig:Variance}, the central map. Another minimum of the width of arrival directions at energies above 55 EeV is revealed in data provided by the TA Collaboration (\cite{TA}, mapped on the right).

To estimate the probability of MWADD under $H_0$ being less than or equal to the observed value, the MC algorithm is used with the number of isotropic events in a set equal to the number of observed EAS events, $N_{obs}$, in the particular energy bin. The exact algorithm of MWADD calculation that was used for the data is then performed on the MC event set.

The procedure is repeated M times to find the fraction of MC event sets where the minimal distribution width is equal or less than the experimental value. This fraction is interpreted as a probability to quote the significance of the anisotropy signal. The number of MC event sets used in simulation and the number of events in energy bins detected in experiments are presented in Table 1.

\begin{table}[t]
\caption{MC simulation results: the probability, $P_0$, of MWADD under $H_0$ to be less than or equal to the observed value. The number of EAS events observed in energy bins, $N_{obs}$, and a sample size, $M$, used in simulation are given for arrays.}
\center\begin{tabular}{rrrr}&&&\\
\hline
Experiment, energy bin, EeV & $N_{obs}$ &    $M$ & $P_0$,\%\\
\hline
           PAO, $E_{WG}>54$ &       231 &  10000 &     57.3\\
            TA, $E_{WG}>55$ &        72 & 100000 &      0.1\\
      Yakutsk, $( 3.1,5.6)$ &       939 &  10000 &     98.5\\
      Yakutsk, $(5.6,10.0)$ &       285 &  10000 &      0.1\\
     Yakutsk, $(10.0,17.7)$ &        95 & 100000 &     14.6\\
     Yakutsk, $(17.7,56.1)$ &        42 & 100000 &     23.5\\
\hline
\end{tabular}\end{table}

The resultant probability for the Yakutsk array data in the energy bin $E_{WG}\in(5.6,10)$ EeV is $P_0=1.15\times 10^{-3}$, which is equivalent to $\sim3.1\sigma$ deviation in the normal distribution terms. However, a penalty factor should be applied to the probability, which is calculated \textit{a posteriori}. Assuming equally possible anisotropy in any of the four energy bins, with comparable deviations, one has a final probability $P=4.6\times 10^{-3}$, equivalent to $\sim2.6\sigma$. Consequently, the null hypothesis can be rejected basing on the Yakutsk Array data at the significance level of $99.5\%$.

\begin{figure}[b]\centering
\includegraphics[width=\columnwidth]{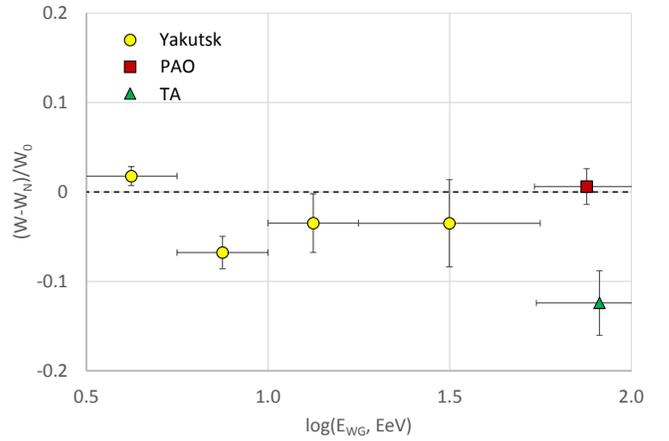}
  \caption{Minimal width of the distribution of arrival directions as a function of CR energy. A normalized difference in the widths of observed and isotropic distributions is derived using the data from EAS arrays.}
\label{Fig:MinWidth}\end{figure}

Observed values of MWADD calculated using available data from PAO \citep{PAOdata} and TA \citep{TA} are shown in comparison with the Yakutsk Array data in Fig. \ref{Fig:MinWidth}. There is no deviation from isotropic expectation in the data from PAO, while TA data exhibit a pronounced deviation of MWADD in the energy range where a `hotspot' was indicated \citep{TA}.

The probability of 72 isotropic EAS events above $E_{WG}>55$ EeV having a MWADD less than that observed by the TA is $P_0=1.3\times10^{-3}$ ($\sim3\sigma$). A penalty factor can be calculated by the TA Collaboration only using all the data observed.

%%%%%%%%%%%%%%%%%%%%%%%%%%%%%%%%%%%%%%%%%%%%%%%%%%%%%%%%%%%%%%%%%%%%%%%%%%%%%%%%%%%%
\subsubsection{Searching for the coordinates of possible CR sources and fraction of the flux}
Our alternative hypothesis, $H_1$, has free parameters to adjust to the observed MWADD: the position of SS in an equatorial system and the fraction of the total CR flux that has arrived from the source. The same iterative procedure is used to calculate the `most probable' parameters yielding the observed distribution width. As an implementation, the MC program mentioned above is adapted to calculate MWADD under $H_1$ with input free parameters.

Fitted parameters are then used to calculate the random dispersion of MWADD for a fixed $N$ equal to the number of detected CRs in a particular energy bin. By varying a parameter, its confidence interval is determined where the resultant deviation of MWADD is within random dispersion limits.

Hypothesis $H_1$ is applied to the Yakutsk Array data in the energy interval $E_{WG}\in(5.6,10)$ EeV and to TA data at $E_{WG}>55$ EeV. The most probable parameters for the Yakutsk Array data are $\alpha_Y=36^0$~$^{+33}_{-30}$,
$\delta_Y=48^0$~$^{+25}_{-17}$, $f_Y=0.11\pm0.08$. The results for TA are $\alpha_{TA}=144^0$~$^{+29}_{-30}$, $\delta_{TA}=42^0$~$^{+23}_{-22}$, $f_{TA}=0.2\pm0.1$.

A hint of the possible source of CRs in the interval $E_{WG}\in(5.6,10)$ EeV, derived from the Yakutsk Array data using the MWADD method in the right ascension circle \citep{MinVar}, is confirmed; the resulting $\alpha_Y$ intervals are within experimental errors.

The coordinates of a hypothesized TA source are in agreement with that of a hotspot revealed by the TA Collaboration \citep{TA}. An additional bonus in our case is an estimation of the probable fraction of CRs attributed to a source.

%%%%%%%%%%%%%%%%%%%%%%%%%%%%%%%%%%%%%%%%%%%%%%%%%%%%%%%%%%%%%%%%%%%%%%%%%%%%%%%%%%%%
\section{Conclusion}
The Yakutsk Array data on arrival directions of CRs above $3.2$ EeV in equatorial coordinates are analyzed using the minimal width of distribution method. A previous hint of large-scale anisotropy in the energy range $5.6<E_{WG}<10$ EeV is confirmed by the enhanced method. The null hypothesis is rejected at the $99.5\%$ significance level.

For comparison, our method of analysis is applied to available data from other giant EAS arrays. PAO data demonstrated no deviation of the MWADD from the isotropic distribution width at energies above 54 EeV. On the contrary, arrival directions of UHECRs detected with TA have a decreased minimal width at $E_{WG}>55$ EeV with a statistical significance of $\sim3\sigma$.

The MWADD method is applied with an alternative hypothesis of a single CR source in a weighed combination with the uniform background. Free parameters of a source are fitted using experimental data in the energy intervals where $H_0$ is ruled out: $\alpha_Y=36^0$~$^{+33}_{-30}$, $\delta_Y=48^0$~$^{+25}_{-17}$, $f_Y=0.11\pm0.08$ for the Yakutsk Array data in the energy interval $E_{WG}\in(5.6,10)$ EeV, and $\alpha_{TA}=144^0$~$^{+29}_{-30}$, $\delta_{TA}=42^0$~$^{+23}_{-22}$, $f_{TA}=0.2\pm0.1$ for TA data at $E_{WG}>55$ EeV.

Although our alternative hypothesis is not a unique explanation of the observed decrease in MWADD, the indicated coordinates and fitted CR fraction of the possible sources may be useful in a future search with enhanced statistics for anisotropy in arrival directions of UHECRs.

The equatorial coordinates of the possible source derived from TA data are in agreement with a hotspot position found by comparison of CR events summed within sky regions \citep{TA}. Our approach is different, being based on the overall distribution width of arrival directions rather than on the excess flux of CRs in a particular angular region.

\acknowledgements
The author is grateful to the Yakutsk Array staff for the data acquisition and analysis. The work is supported in part by the Russian Academy of Sciences (Program 10.2) and RFBR (grants 11-02-00158 and 13-02-12036).

\appendix
\section{}
Our objectives in the paper are a circular uniform distribution on the unit sphere $S^2$ in $\mathbb{R}^3$ and, as antithesis, a point source of CRs. For simplicity, all considerations in the Appendix will be illustrated on a circle, where $\psi_0$ is the point source position.

The most common way in directional statistics is to use $n$-th moments of a distribution defined as $m_n=\frac{1}{N}\sum_{i=1}^N\exp(in\psi_i)$, where $N$ is a number of data points at $\psi_i$, with the mean angle $\overline{\psi}=Arg(m_1)$ and the circular variance $S=1-|m_1|$. However, there are other measures of the distribution moments in use. For instance, we are using in the paper a measure of dispersion of angles $\frac{1}{N}\sum_{i=1}^N(\pi-|\pi-|\psi_i-\psi_0||)$ considered by \cite{Mardia}. In this approach the mean angle is that where dispersion reaches the minimum (for distributions under consideration the median coincides with the mean).

The MWADD method is used in the paper to locate the mean and to test for uniformity of directions against alternative hypothesis with a point source. In Section 3.2, the method is compared to the Rayleigh test known as one of the most powerful tests for unimodal data \citep{Jupp}.

%%%%%%%%%%%%%%%%%%%%%%%%%%%%%%%%%%%%%%%%%%%%%%%%%%%%%%%%%%%%%%%%%%%%%%%%%%%%%%%%%%%%

\end{document}